# The dispersion of research performance within and between universities as a potential indicator of the competitive intensity in higher education systems[1]

*Giovanni Abramo*[a,b,*], *Tindaro Cicero*[b], *Ciriaco Andrea D'Angelo*[b]

[a] Institute for System Analysis and Computer Science (IASI-CNR)
National Research Council of Italy

[b] Laboratory for Studies of Research and Technology Transfer
School of Engineering, Department of Management
University of Rome "Tor Vergata"

**Abstract**

Higher education systems in competitive environments generally present top universities, that are able to attract top scientists, top students and public and private financing, with notable socio-economic benefits in their region. The same does not hold true for non-competitive systems. In this study we will measure the dispersion of research performance within and between universities in the Italian university system, typically non-competitive. We will also investigate the level of correlation that occurs between performance in research and its dispersion in universities. The findings may represent a first benchmark for similar studies in other nations. Furthermore, they lead to policy indications, questioning the effectiveness of selective funding of universities based on national research assessment exercises. The field of observation is composed of all Italian universities active in the hard sciences. Research performance will be evaluated using a bibliometric approach, through publications indexed in the Web of Science between 2004 and 2008.

\* **Corresponding author**, Dipartimento di Ingegneria dell'Impresa, Università degli Studi di Roma "Tor Vergata", Via del Politecnico 1, 00133 Rome - ITALY, tel. +39 06 72597362, abramo@disp.uniroma2.it

# 1. Introduction

In every field of human activity there are certain individuals who distinguish themselves by particular competencies, attitudes and interests, generally translating into outstanding levels of performance. Similarly, in every economic sector there are certain organizations, public or private, for-profit or not-for-profit, which consistently demonstrate top performance, in turn leading to a reputation which is likely to become a distinctive competence of the organization. These organizations generally excel at every link in the chain of value, thanks to quality of their personnel and the strategic and organizational capacity of their management. The fundamental key to their long-term success is a capacity to attract, develop and retain talent, meaning the best representatives of the work force in each function. Research organizations (universities and institutions), whether public or private, will not escape this rule as long as they operate in competitive environments. The level of domestic competition in higher education is determined by several contextual elements, from cultural practices to political legitimization of the system. One of the fundamental determinants of competitive intensity is undoubtedly the typology of funding and related incentives, at macro, meso and micro levels. In countries where government funding of universities represents a large share of total budget and is allocated with the intention of ensuring sufficiency of resources, competition among universities will likely be low in intensity. On the contrary, in nations where extra-governmental financing is significant and the public allocation is essentially based on merit, the environment is likely to be more competitive (Geuna and Martin, 2003). Over time, competition should lead to the development of distinctive competencies and to a subsequent competitive advantage of some organizations over all the others, meaning that in competitive environments it is possible to observe more marked performance differences among universities, and a mapping of the higher education industry into strategic groups.

Auranen and Nieminen (2010) classified the higher education systems of such countries as Germany, Sweden and Denmark (the present authors would also add Italy) as non-competitive. In these nations, the distinction of the different levels of excellence among universities is not immediate. On the contrary, in countries such as the United States, United Kingdom and Australia, which fall in the competitive category, few would have difficulty in identifying the best research universities. These universities are in competition with one another to bring in, from at home and abroad, the best researchers and teaching professors, the best technical-administrative personnel, best students, donations and public and private financing for research. Their reputations and competitive advantages are reflected as much in the salary levels and status that they succeed in providing as in the willingness of students to pay tuition fees that are well over average. Local and national governments have every interest in developing and cultivating champion institutions. In fact, the social and economic benefits for the nation and home region of a prestigious university have been soundly demonstrated (Rosemberg and Nelson, 1994, Pressman et al., 1995; Fritsch and Slavtchev, 2006). Governments should thus provide the conditions for development of competitive environments, leading to continuous improvement in the entire higher education system, and the emergence of top universities.

It is no accident that, in recent years, an increasing number of nations have begun regular national exercises for research evaluation, permitting the allocation of funds on the basis of performance criteria and the stimulation of greater research productivity.

These exercises are generally based on the peer-review method, implying the evaluation of only a share of the entire research output of each university. Rankings for the universities are drawn up on the basis of the average quality of the research products as submitted by the institutions, and not on productivity. Performances of universities are compared at discipline level and not at individual level.

If the objective of government is to stimulate competition among universities, and so lead to the development of top universities that can compete internationally and produce the relevant socio-economic benefits, then one of the indicators to monitor over time, other than performance itself, is the degree of concentration of performance within universities. The less research performance is dispersed within universities then the greater is the probability of having top universities that are able to compete at the international level. This would indicate that the competitive environment has led to the formation of a system of universities with marked differences in performance, and thus the concentration of top scientists in top universities. In the American system of higher education, which appears to be among the most competitive, presenting a significant number of recognized universities able to attract resources and students from around the world, we expect that the average performance of researchers in these top universities would be very high, but that the variability, and thus the concentration, would be low. Universities of average level would be expected to show an average performance, perhaps with points of excellence in one or two disciplines, but again with a quite low concentration. To achieve balanced regional development, an additional policy objective could be that the top universities be distributed quite uniformly throughout the nation.

To date, the literature does not offer detailed analyses of the concentration of performance in university systems. In 1974, Allison and Stewart detected a highly skewed distribution of productivity (publications and citations) among 2,172 scientists working in US academic departments. The authors pointed out the presence of accumulative advantage, so that "because of feedback through recognition and resources, highly productive scientists maintain or increase their productivity, while scientists who produce very little produce even less later on". Daniel and Fisch (1990), in their review on the history of research performance evaluation in German universities, stressed the importance to take into account concentration of performance: in their opinion, when the department is the unit of analysis, the evaluator "should distinguish between those with 'collective strength' and those with 'individual strength' by computing some measure of concentration". However, the paper provides no empirical evidence about concentration: the authors make only reference to one German leading psychology department and recall fundamental regularities of the distribution of scientific productivity, already pointed out in pioneering works of Lotka (1926) and De Solla Price (1971). The lack of studies on concentration of research performance is easily understandable: realizing such analysis requires the measurement of performance at the level of single scientists, which can be hardly achieved on large populations. Meanwhile, except for Abramo and D'Angelo (2011), the bibliometric exercises available in the literature have been limited to single institutions or disciplines (Kalaitzidakis et al. 2003; Macri and Dipendra, 2006;

Gonçalves et al., 2009; Costas, 2010), and have never entered into the comparative analysis of the degree of concentration of performance in the institutions examined.

What we propose for this work is exactly this measurement of the degree of concentration of performance, in the hard sciences for Italian universities. We will subsequently compare the variability of performance within universities to that between universities. Finally, we will examine the level of correlation between degree of concentration and average level of performance. Through the findings of our analyses, we aim at providing policy indications on the risks of assigning public funds to universities on the basis of the outcomes of national research assessment exercises. Furthermore, we hope to stimulate similar analyses in other national contexts in order to demonstrate the assumption that non-competitive higher-education systems differ from competitive ones in terms of concentration of performance within and between universities.

The next section of the paper presents a summary description of the Italian university system. Section 3 describes the methodology adopted, the dataset and indicators used. Section 4 presents the results, with discussion. The last section presents the pertinent conclusions from the work, with the authors' comments.

**2. The Italian university system**

In Italy, the Ministry of Education, Universities and Research (MIUR) recognizes a total of 95 universities, with the authority to issue legally-recognized degrees. With only rare exceptions these are public universities. These universities are largely financed through non-competitive allocation from the MIUR, although this share of income is decreasing, as seen in the reduction from 61.5% in 2001 to 55.5% in 2007 (MIUR, 2010). In keeping with the "welfare state" view that higher education is a public good which should be made accessible to all, from the wealthy to the less advantaged, the government imposes a price-cap for tuition fees. The effective fees varies from student to student, since fees are calculated according to family income. However they are very low, roughly 1,000 Euro per academic year[2], and as such provide 12.5% of total university income. Further financing from the MIUR, for research projects on a competitive basis[3], represents an additional 8.8% of income. Other public and private financing for research projects, obtained on a competitive basis, adds a further 17.0% of total income. Income deriving from technological transfer is negligible, given the very limited practice of Italian universities to carry out patenting and licensing (Abramo, 2007). Donations to universities are not a feature of the Italian cultural tradition and are also negligible. Up to 2009, the core funding by government was input oriented, meaning that it was distributed to universities in a manner intended to satisfy the needs for resources of each and all, in function of their size and activities. It was only following the first national research evaluation exercise (VTR), conducted between 2004 and 2006, that a minimal share, equivalent to 7% of MIUR financing[4], was attributed in function of the research evaluation and of teaching quality. All new personnel enter the university system through public examinations, and career advancement also requires such public examinations. Salaries are regulated at the nationally centralized level and are calculated according to role (administrative, technical, or professorial), rank within role (for example: assistant, associate or full professor), and seniority. No part of the salary for professors is related to merit: wages increase annually according to parameters set by government. All professors are contractually obligated to carry out research, thus there is no development of research and teaching universities. The whole of these conditions has created an environment and culture that is completely non-competitive.

---

[2] As a comparison, annual tuition fees at Oxford University are 4,000 Euro; at Stanford University annual fees are approximately 30,500 Euro

[3] In the Italian university sphere there is great skepticism that the financing of research projects by the public sector is truly determined on a merit basis, due to a history of strongly-rooted favoritism practices in the environment and perceptions of poor effectiveness in procedures for project evaluation.

In this completely non-competitive context, lacking incentive for improvement, we would expect that research performance among Italian universities will be more or less similar. Potential differences would be due to external factors, such as: i) economic rents due to location, in favor of universities situated in areas of high intensity for private research, where geographic proximity effect (Jaffe, 1989; Lang, 2005; Coronado and Acosta, 2005) would permit greater opportunities for private financing and stimuli; and ii) corruption in the boards of examiners for national competitions, a phenomenon that a succession of governments addressed through reform (Zagaria, 2007; Perotti, 2008), with varying degrees of earnestness but always without success, meaning that the winners of competitions are not always the most deserving candidates.

Given the context, we would expect an assessment of research performance in Italian universities to reveal: i) a distribution of performance within universities which is generally highly concentrated and similar to that of the entire research population; and ii) a variability of average performance among universities that is less than the variability of performance of the individual scientists within each university. This is what we will attempt to verify, after having presented the methodology with which we conduct the study.

## 3. Methodology

National-scale comparisons of research performance at the individual level is not an easy task. Bibliometric techniques are apt to the scope, but few formidable obstacles need to be overcome, namely: i) the reconciliation of the authors' affiliations and the attribution of research output to its real author, which requires an in depth knowledge of the research system of the country under observation; and ii) the reduction of performance measurement distortions due to the varying publication and citation intensities of the different fields of research.

The research performance will be observed over the period 2004-2008. The field of observation (see Section 3.1) is composed of all researchers in the hard sciences in Italian universities, as indexed in an MIUR database[5]. Research performance will be measured using bibliometric techniques, as presented in Section 3.2, essentially using indicators of productivity, i.e. the number of publications produced; and impact, i.e. the number of field standardized citations. Limiting the field of investigation to the hard sciences, the literature certainly gives ample justification for the choice of considering: i) scientific publications indexed in Web of Science™ (WoS) as proxy of overall research output (Moed et al., 2004); and ii) citations as proxy of impact on scientific advancement, notwithstanding the possible distortions inherent in these indicators (Glanzel, 2008). The degree of concentration of performance by the researchers in the various disciplines will essentially be measured using the Gini coefficient[6] adjusted for small samples (Deltas, 2003), but also through the ratio of the cumulative bottom 40% to top 20% of performances.

---

[4] Since MIUR financing composes 55.5% of the total, the share that is distributed on the basis of the VTR represents 3.9% of total income.
[5] http://cercauniversita.cineca.it/php5/docenti/cerca.php

**3.1 Dataset**

The dataset used for the analysis is composed of the raw data, acquired from Thomson Reuters, of the publications (articles, reviews and conference proceedings) listed in the Italian National Country Report, extracted from the WoS. Based on this data, using a complex algorithm for reconciling the authors' affiliation and disambiguating the true author identities, each publication is attributed to the university scientist(s) who produced it A thorough description of the algorithm may be found in D'Angelo et al. (2011).

The Italian academic system provides that each researcher belongs to a specific Scientific Disciplinary Sector, or SDS. Each SDS is in turn part of a University Disciplinary Area, or UDA. The hard sciences are gathered in nine UDAs[7] and 205 SDSs. To render greater significance, the field of observation was limited to those SDSs where at least 50% of member scientists produced at least one publication in the period 2004-208. There are 184 such SDSs[8]. Over the period examined, these 184 SDSs contained an average of 39,512 scientists distributed in 77 universities (Table 1).

| UDA | SDSs | Universities | Research staff |
|---|---|---|---|
| Agricultural and veterinary sciences | 28 | 48 | 3,153 |
| Biology | 19 | 66 | 5,785 |
| Chemistry | 12 | 59 | 3,607 |
| Civil Engineering and architecture | 7 | 49 | 1,455 |
| Earth sciences | 12 | 48 | 1,439 |
| Industrial and information engineering | 42 | 68 | 5,489 |
| Mathematics and computer sciences | 9 | 64 | 3,515 |
| Medicine | 47 | 55 | 12,196 |
| Physics | 8 | 61 | 2,873 |
| Total | 184 | 77 | 39,512 |

*Table 1: Research staff and number of SDSs per Italian UDA; data 2004-2008.*

---

[6] Gini coefficient is the most commonly used measure of inequality. It varies between 0, which reflects complete equality and 1, which indicates complete inequality (one person has all the measure, all others have none).

[7] Mathematics and computer sciences; physics; chemistry; earth sciences; biology; medicine; agricultural and veterinary sciences; civil engineering and architecture; industrial and information engineering.

[8] The complete list is provided on:
http://www.disp.uniroma2.it/laboratorioRTT/TESTI/Indicators/ssd2.html

**3.2 Indicators**

For measurement of research performance we have formulated two indicators, with differences in calculation according to the level of analysis: individual researchers or entire universities. The indicators are named productivity (*P*) and scientific strength (*SS*). At the level of individual researchers *P* is calculated as the sum of publications of an author, each divided by the number of co-authors[9].

The classification of university research staff into SDSs helps reduce productivity measurement distortions. In fact, variation of publication intensity of subfields within the same SDS is expected to be small enough to make productivity comparisons equitable. Abramo et al. (2008) noted though that a number of researchers within the same SDS publish in more than one WoS subject category[10]. This calls for a field (subject category) standardization, when comparing impact of researchers within the same SDS, to account for the varying citation intensity of different subject categories. The impact indicator named scientific strength (SS) is defined as the sum of the standardized citations to publications divided by the number of co-authors of each publication[11]. Citations of a publication are standardized dividing them by the median of citations[12] of all Italian publications of the same year and WoS subject category[13].

Values of both indicators are averaged on the number of years in which each scientist was officially on staff in Italian universities over the five years under examination, so to have an average annual performance.

For the analysis of performance at university level, the calculation of performance at SDS level is required first. At SDS level, in formulae, productivity (*P*) and scientific strength (*SS*) of university *i* in the SDS *s* are, respectively:

$$P_{i,s} = \frac{1}{RS_{i,s}} \cdot \sum_{j=1}^{N_s} n_{j,i,s}$$

$$SS_{i,s} = \frac{1}{RS_{i,s}} \cdot \sum_{j=1}^{N_s} \frac{C_j}{\bar{C}_j} \cdot n_{j,i,s}$$

With:

$n_{j,i,s}$ = fraction of authors of university *i* in SDS *s* on total co-authors of publication *j*, (considering, if publication *j* falls in life science subject categories, the position of each author in the list and the character of the co-authorship, intra-mural or extra-mural).

$N_s$ = total number of publications in SDS *s*

$RS_{i,s}$ = average research staff of university *i* in SDS *s*, in the observed period

$C_j$ = number of citations received by publication *j*

---

[9] In the case of the life sciences, different weights have been given to each co-author according to his/her position in the list and the character of the co-authorship (intra-mural or extra-mural).
[10] Journals indexed in the WoS Science Citation Index are classified into 182 subject categories.
[11] See above footnote.
[12] Observed as of 30/06/2009.
[13] The choice to standardize citations with respect to the median value (rather than to the average, as is frequently observed in the literature) is justified by the fact that the distribution of citations is highly skewed in almost all subject categories (Lundberg, 2007).

$\bar{C}_j$ = median of citations received by all Italian publications of the same year and subject category of publication $j$

Optionally, the performance may be calculated at UDA level, aggregating performance data of each SDS within the UDA. To account for: i) varying publication and citation intensities of different SDSs and ii) differing representativeness, in terms of research staff, of the SDSs present in each UDA, data are conveniently: i) standardized and ii) weighted. At UDA level, in formulae, productivity ($P$) and scientific strength ($SS$) of university $i$ in the UDA $u$ are, respectively

$$P_{i,u} = \frac{1}{RS_{i,u}} \cdot \sum_{s=1}^{N_u} \frac{P_{i,s}}{\bar{P}_s} \cdot RS_{i,s}$$

$$SS_{i,u} = \frac{1}{RS_{i,u}} \cdot \sum_{s=1}^{N_u} \frac{SS_{i,s}}{\overline{SS}_s} \cdot RS_{i,s}$$

With:
$RS_{i,u}$ = average research staff of university $i$ in UDA $u$, in the observed period
$N_u$ = number of SDSs in the UDA $u$
$\bar{P}_s$ = average value of productivity of all universities active in SDS $s$
$\overline{SS}_s$ = average value of scientific strength of all universities active in SDS $s$

The performance at university level is obtained in a similar way, aggregating standardized and weighted data of all SDSs in which a university is active.

**4. Results**

The results from the analyses are presented in the series of following subsections. First we present the analysis of concentration of performance by university researchers in each SDS and UDA at the overall Italian level, to observe potential differences among SDSs and UDAs. Next is an analysis at the level of single universities, to position each one in terms of degree of concentration of performance with respect to the overall Italian datum. Then we present results from the comparison between the variability of performance by researchers in the same discipline and university and the variability of average performance between universities. Finally we present the results from analysis of the correlation between average performance of each university and its concentration of performance, to understand if greater or lesser concentration can in any way be associated with level of performance.

**4.1 Concentration of performance of scientists**

In this subsection we analyze the distribution of performance by the individual scientists of the entire national dataset. We first quantify the degree of concentration at the level of the single SDSs and then of whole UDAs, to understand if and how the scientific performance of researchers is heterogeneous and if there are significant differences among SDSs and among UDAs.

We show the procedure used, presenting the example of SDS FIS/01 (Experimental physics). Over the five years from 2004-2008 this sector included 1,117 researchers distributed amongst 53 universities. For 133 of these (11.9% of the total), there were no publications registered over the period (P=0). Of the remaining 984, 60 (an additional 5.4%) show a null SS, having produced works that received no citations. The degree of concentration, measured using the adjusted Gini coefficient (in the following named Gini coefficient, for simplicity's sake), is 0.587 for P and 0.690 for SS. Figure 1 presents the Lorenz curves. It can be seen that the curves diverge notably from the diagonal, indicating equidistribution of performance. The ratio to the cumulative performance of the bottom 40% to the top 20% of all the scientists, without considering university affiliation, is 8.3% for P and 2.9% for SS.

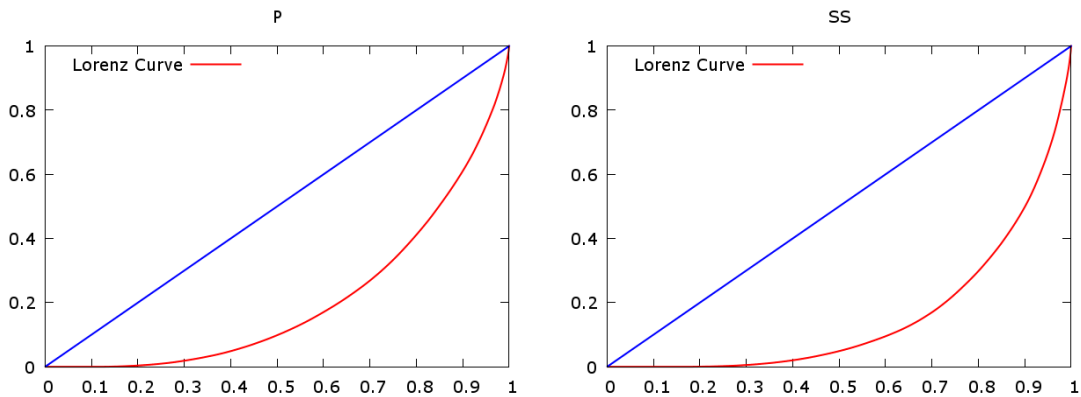

*Figure 1: Lorenz curves for distributions of P and SS of university researchers in the SDS "Experimental physics, data 2004-2008.*

The next steps are the analysis of distribution of performance and the calculation of the relative Gini coefficients for all 184 SDSs. The descriptive statistics for the Gini coefficient data are depicted by the box plots for P and SS of Figure 2. The box represents the interquartile range (IQR), meaning the interval of observations between the first and third quartile (respectively indicated by the lower and upper borders of the box). The line that divides the box in two parts represents the median. The whiskers below and above the box represent the lowest datum still within 1.5 IQR of the lower quartile, and the highest datum still within 1.5 IQR of the upper quartile. It can be seen that the distribution of coefficients of concentration for P of the SDSs is symmetrical relative to the median (0.581), going from a minimum of 0.297. to a maximum of 0.828. The distribution for the concentration of SS presents a median value of 0.710 and a still more pronounced negative asymmetry. The minimum value is 0.156, the maximum is 0.996, both anomalous values compared to the totality of the data.

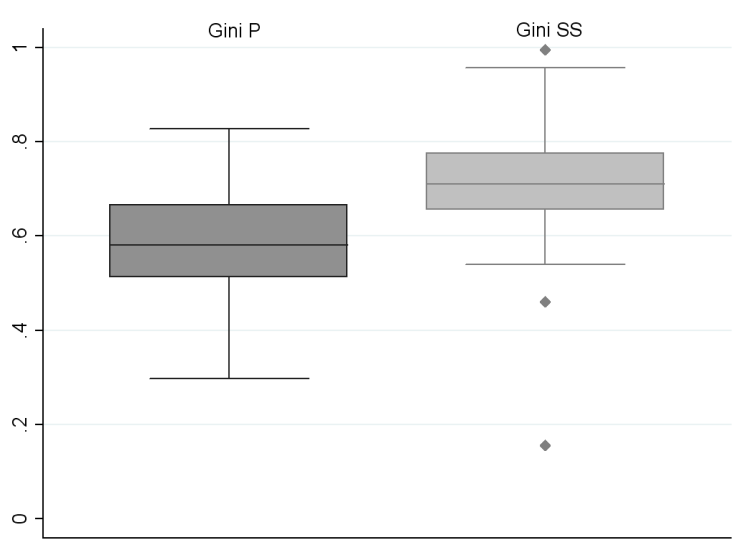

*Figure 2: Box plots of Gini coefficients for P and SS in the 184 SDSs*

The individual coefficients of concentration measured for SDSs are now weighted for the share of scientists on staff in the SDS relative to the total for the UDA to which it belongs, thus obtaining an average value of concentration at the level of UDA for both measures of performance. Table 2 presents the descriptive statistics for the distribution of Gini coefficient by UDA.

| UDA | n. SDS | Weighted mean of Gini coeff. | SDS min | max | median | std dev. |
|---|---|---|---|---|---|---|
| Agricultural and veterinary sciences | 28 | 0.560 | 0.424 | 0.726 | 0.554 | 0.085 |
| Biology | 19 | 0.565 | 0.472 | 0.725 | 0.548 | 0.076 |
| Chemistry | 12 | 0.485 | 0.297 | 0.572 | 0.465 | 0.070 |
| Civil Engineering and Architecture | 7 | 0.625 | 0.568 | 0.660 | 0.631 | 0.033 |
| Earth sciences | 12 | 0.556 | 0.455 | 0.689 | 0.540 | 0.069 |
| Industrial and Information engineering | 42 | 0.536 | 0.417 | 0.828 | 0.551 | 0.101 |
| Mathematics and computer sciences | 9 | 0.570 | 0.456 | 0.682 | 0.573 | 0.080 |
| Medicine | 47 | 0.670 | 0.462 | 0.772 | 0.674 | 0.070 |
| Physics | 8 | 0.548 | 0.469 | 0.777 | 0.539 | 0.093 |

*Table 2: Descriptive statistics of Gini coefficients of P for each UDA*

For SS, the different areas show a value of concentration that averages higher.

| UDA | n. SDS | Weighted mean of Gini coeff. | SDS min | max | median | std dev. |
|---|---|---|---|---|---|---|
| Agricultural and veterinary sciences | 28 | 0.713 | 0.574 | 0.882 | 0.709 | 0.087 |
| Biology | 19 | 0.682 | 0.559 | 0.827 | 0.684 | 0.063 |
| Chemistry | 12 | 0.611 | 0.156 | 0.662 | 0.597 | 0.138 |
| Civil Engineering and architecture | 7 | 0.769 | 0.677 | 0.819 | 0.768 | 0.046 |
| Earth sciences | 12 | 0.679 | 0.578 | 0.818 | 0.653 | 0.070 |
| Industrial and Information engineering | 42 | 0.704 | 0.540 | 0.996 | 0.712 | 0.101 |
| Mathematics and computer sciences | 9 | 0.724 | 0.601 | 0.803 | 0.730 | 0.062 |
| Medicine | 47 | 0.753 | 0.604 | 0.861 | 0.756 | 0.058 |
| Physics | 8 | 0.654 | 0.586 | 0.866 | 0.661 | 0.086 |

*Table 3: Descriptive statistics of Gini coefficients of SS for each UDA*

The Chemistry area is notable for the lower concentration values for its SDSs, in both measures of performance.

The analyses permit the conclusions that the level of concentration of performance in SDSs and UDAs is high for P and very high for SS, and that there are no substantial differences between the UDAs, in spite of the occurrence of a few outlier SDSs.

**4.2 Concentration of performance within universities**

In this section we analyze the distribution of performance within individual universities and then compare the institutions, in terms of degree of concentration, with what we saw in the previous section, where the distributions referred to all the national researchers of a given SDS without considering the institutions to which they belong. As an example, referring to SS, Figure 3 presents the dot plot of Gini coefficients for universities with researchers in SDS FIS/01. For significance reasons, analysis was limited to universities with at least 5 researchers in the SDS, numbering 42 in total.

For these universities and this SDS, the degree of concentration of SS is demonstrated as falling around the national value, as represented by the horizontal line in Figure 3. Apart from a few exceptions, the greater part of universities show a Gini coefficient between 0.50 and 0.80.

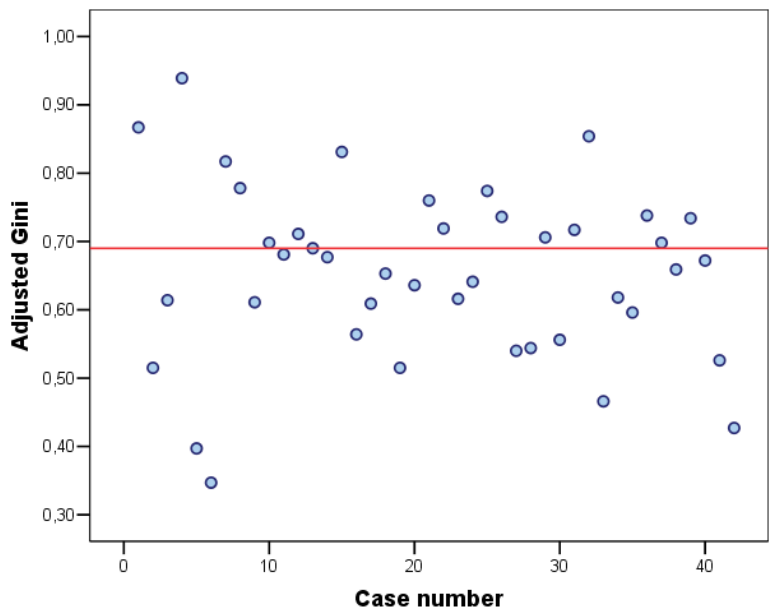

*Figure 3: Dot plot of Gini coefficients for SS in universities with researchers in the "Experimental physics" SDS.*

Table 4 presents the details for the measurements. In addition to the Gini coefficient and ratio of bottom 40% to top 20%, the table shows the values for SS. The universities were subdivided into 4 classes in function of their level of concentration: i) very high inequality, Gini coefficient above 0.70; ii) high inequality, Gini 0.60-0.70; iii) moderate inequality, Gini 0.50-0.60; iv) low inequality, Gini under 0.50. Each class was then subdivided into two sections, separating universities with an SS above the average from those with a below average SS. The average value of SS, for the 42 universities active in FIS/01, is 0.445. There are a total of only four universities in the classes "low inequality" and eight universities in "moderate inequality". The greater part of the universities are situated in classes "high" and "very high" inequality. In this SDS it seems that there is an inverse relationship between the degree of concentration of performance and the value of performance within each university. In fact, proceeding from the low through to the very high inequality class, the number of universities above the average performance value for the sector tends to diminish.

|  | Very High inequality Gini 0.70 ↑ | | | | High inequality Gini 0.60-0.70 | | | | Moderate inequality Gini 0.50 – 0.60 | | | | Low inequality Gini 0.50↓ | | | |
|---|---|---|---|---|---|---|---|---|---|---|---|---|---|---|---|---|
|  | University | SS | Ratio | GINI | University | SS | Ratio | GINI | University | SS | Ratio | GINI | University | SS | Ratio | GINI |
| *Above mean SS* | Brescia | 1.152 | 0.065 | 0.711 | Milan | 0.757 | 0.024 | 0.653 | Milan Polytechnic | 1.109 | 0.140 | 0.515 | Venice - "Ca' Foscari" | 0.623 | 0.501 | 0.347 |
|  | Salento | 1.002 | 0.011 | 0.778 | Turin Polytechnic | 0.611 | 0.090 | 0.614 | Trent | 0.956 | 0.164 | 0.526 | Marche Polytechnic | 0.611 | 0.393 | 0.427 |
|  | Palermo | 0.860 | 0.000 | 0.719 | Milan Bicocca | 0.566 | 0.06 | 0.609 | Modena and Reggio E. | 0.757 | 0.153 | 0.515 |  |  |  |  |
|  | Piedmont Orient.Av. | 0.514 | 0.038 | 0.817 |  |  |  |  | Messina | 0.489 | 0.146 | 0.564 |  |  |  |  |
|  | Rome "La Sapienza" | 0.498 | 0.019 | 0.736 |  |  |  |  |  |  |  |  |  |  |  |  |
|  | Ferrara | 0.458 | 0.025 | 0.734 |  |  |  |  |  |  |  |  |  |  |  |  |
| *Below mean SS* | Superior Normal School | 0.395 | 0.000 | 0.939 | Naples "Federico II" | 0.425 | 0.070 | 0.636 | Calabria | 0.365 | 0.093 | 0.596 | II Napoli | 0.418 | 0.054 | 0.397 |
|  | Turin | 0.373 | 0.029 | 0.706 | Florence | 0.419 | 0.049 | 0.677 | Rome "Tor Vergata" | 0.290 | 0.162 | 0.540 | Urbino "Carlo Bo" | 0.404 | 0.135 | 0.466 |
|  | Padua | 0.351 | 0.007 | 0.760 | Parma | 0.415 | 0.033 | 0.616 | Salerno | 0.265 | 0.156 | 0.526 |  |  |  |  |
|  | Trieste | 0.305 | 0.010 | 0.717 | Aquila | 0.404 | 0.031 | 0.698 | Siena | 0.179 | 0.165 | 0.544 |  |  |  |  |
|  | Genoa | 0.268 | 0.002 | 0.831 | Camerino | 0.392 | 0.009 | 0.698 |  |  |  |  |  |  |  |  |
|  | Bari Polytechnic | 0.224 | 0.000 | 0.867 | Basilicata | 0.335 | 0.024 | 0.611 |  |  |  |  |  |  |  |  |
|  | Perugia | 0.213 | 0.003 | 0.774 | Rome "Tre" | 0.314 | 0.008 | 0.618 |  |  |  |  |  |  |  |  |
|  | Bologna | 0.201 | 0.040 | 0.738 | Catania | 0.305 | 0.054 | 0.659 |  |  |  |  |  |  |  |  |
|  | Udine | 0.135 | 0.010 | 0.854 | Cagliari | 0.257 | 0.006 | 0.690 |  |  |  |  |  |  |  |  |
|  |  |  |  |  | Pisa | 0.244 | 0.056 | 0.672 |  |  |  |  |  |  |  |  |
|  |  |  |  |  | Pavia | 0.228 | 0.078 | 0.641 |  |  |  |  |  |  |  |  |
|  |  |  |  |  | Bari | 0.203 | 0.056 | 0.681 |  |  |  |  |  |  |  |  |

*Table 4: Scientific strength and concentration values of Italian universities in the SDS "Experimental physics".*

As indicated in the introduction to this article, in a context of intense competition we would expect to find excellent organizations with top scientists, and vice versa. Intense competition would lead to greater equidistribution of performance within the universities. On the base of this hypothesis, the authors constructed a theoretical scenario in which the researchers of the national SDS are divided into groups of equal numbers (1,087/42) within the 42 universities under examination, in function of their values of scientific strength. The first university is ideally assigned the group with the highest values of scientific strength, and so on, down to the 42nd university with the research staff of the lowest performance. Recalculating the Gini coefficients for the new hypothetical distribution, we obtain very low values, as expected (Figure 4). Except for two outliers, all the universities have Gini coefficients between 0 and 0.10, values that are very much lower than the national aggregate represented by the horizontal line in Figure 4.

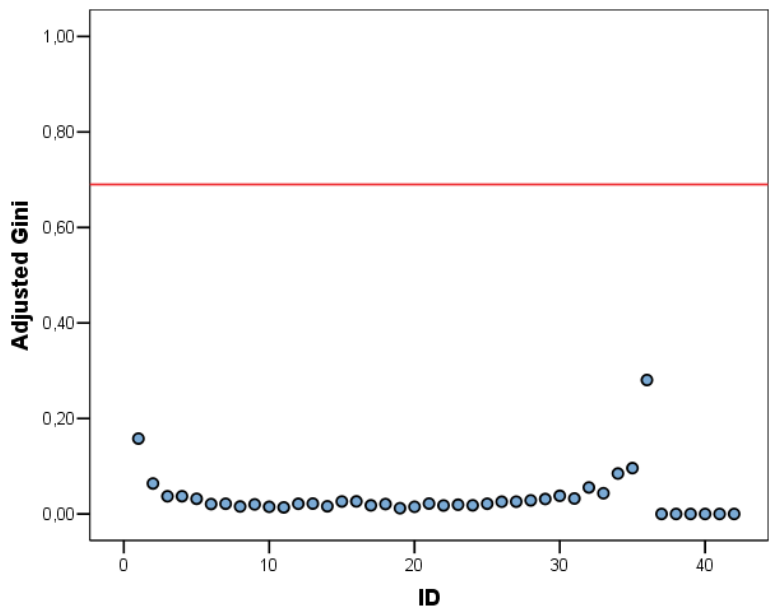

*Figure 4: Dot plot of Gini coefficients for theoretical universities with researchers of the Experimental physics SDS divided amongst them.*

For each SDS, one can measure the divergence of Gini coefficient for distribution of performance from that of the theoretical case. The greater is the shift, the greater is the distance from the "optimal" distribution of research staff among the organizations.

The situation illustrated for this particular SDS does not differ from that of other SDSs in the Physics UDA or those of other UDAs. Table 5 presents descriptive statistics for Gini coefficient of P in each university, by UDA. The concentration of P in the universities varies between 0.151 in Agricultural and veterinary sciences and 1.000 (value registered in 7 out of 9 UDAs). All the UDAs show a median value of concentration above 0.40. Table 6 refers to SS: 7 out of 9 UDAs show at least one university with coefficient of concentration equal to one (third column). The university with the lowest level of concentration occurs in Industrial and information engineering (0.164). Chemistry shows a distribution of Gini coefficients with the lowest values for median (0.547) and maximum (0.844).

|  | Universities | | | |
| --- | --- | --- | --- | --- |
| UDA | min | max | median | std dev. |
| Agricultural and veterinary sciences | 0.151 | 1.000 | 0.536 | 0.166 |
| Biology | 0.166 | 1.000 | 0.543 | 0.153 |
| Chemistry | 0.162 | 0.712 | 0.443 | 0.107 |
| Civil Engineering and architecture | 0.186 | 1.000 | 0.623 | 0.152 |
| Earth sciences | 0.179 | 1.000 | 0.528 | 0.166 |
| Industrial and Information engineering | 0.126 | 1.000 | 0.505 | 0.161 |
| Mathematics and computer sciences | 0.240 | 1.000 | 0.564 | 0.146 |
| Medicine | 0.257 | 1.000 | 0.653 | 0.143 |
| Physics | 0.237 | 0.927 | 0.536 | 0.136 |

*Table 5: Descriptive statistics for Gini coefficient of P for the universities active in each UDA*

|  | Universities | | | |
| --- | --- | --- | --- | --- |
| UDA | min | max | median | std dev. |
| Agricultural and veterinary sciences | 0.181 | 1.000 | 0.669 | 0.168 |
| Biology | 0.222 | 1.000 | 0.646 | 0.152 |
| Chemistry | 0.195 | 0.844 | 0.547 | 0.123 |
| Civil Engineering and architecture | 0.339 | 1.000 | 0.770 | 0.151 |
| Earth sciences | 0.167 | 1.000 | 0.661 | 0.185 |
| Industrial and Information engineering | 0.164 | 1.000 | 0.661 | 0.164 |
| Mathematics and computer sciences | 0.378 | 1.000 | 0.711 | 0.132 |
| Medicine | 0.257 | 1.000 | 0.727 | 0.136 |
| Physics | 0.279 | 0.939 | 0.618 | 0.137 |

*Table 6: Descriptive statistics for Gini coefficient of SS for the universities active in each UDA*

In summary, the distribution of performance within universities is generally highly concentrated. The true situation is very far from the theoretical and optimal case, in which all universities have an almost equidistributed performance. This confirms the expected effects of a non-competitive environment, where in the absence of top universities able to attract the best researchers, these distribute more or less uniformly among the institutions, giving a situation of high concentration of performance.

**4.3 Comparison of performance variability within and between universities**

In a competitive environment we expect that the variability of performance by researchers within universities will be less than that between universities. The more the environment is competitive, the more the difference should be. In a non-competitive environment we expect the opposite.

This section presents the results of the analysis of the variability of performance within universities and between universities in Italy. As an example, we provide the analyses for P in the Physics UDA and for SS in Civil engineering and architecture. For the variability measure we use the coefficient of variation expressed as percentage, so as to compare the variability between distributions with different average values.

In Physics, the level of variability of performance within individual universities, is greater than the variability between them. In Figure 5, the horizontal line represents the value of variability among universities, which is situated under all the individual points, which represent individual universities. The coefficients of variation within universities range from 43% to 162%, while the value between them is 33%.

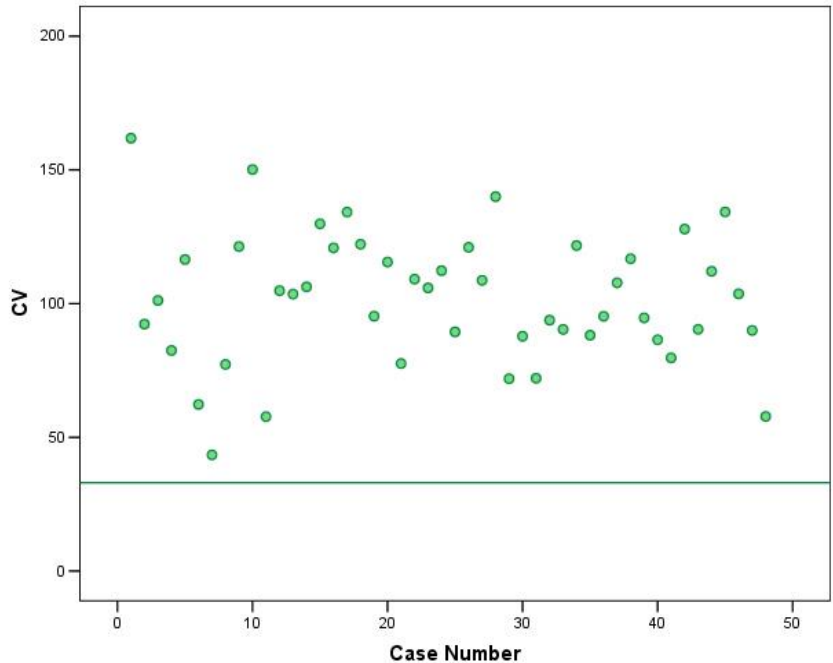

*Figure 5: Coefficients of variation for P within Italian universities (points) in Physics compared with the coefficient between universities (line).*

The situation described for Physics is repeated in all the other UDAs. Table 7 presents descriptive statistics for distributions of the coefficients of variation within the universities in each UDA. Comparing with the variation between universities, we see that, apart from the Civil engineering and architecture UDA, there is no UDA where a university presents variability for performance that is less than that between. The highest median for variability is seen in Medicine (143%) and the lowest is in Chemistry (89%).

|  | $C_V$ within | | | $C_V$ between |
|---|---|---|---|---|
| UDA | min | max | median | |
| Agricultural and veterinary sciences | 35% | 165% | 110% | 32% |
| Biology | 49% | 237% | 110% | 35% |
| Chemistry | 57% | 164% | 89% | 22% |
| Civil engineering and architecture | 57% | 186% | 123% | 59% |
| Earth sciences | 53% | 163% | 98% | 37% |
| Industrial and Information engineering | 53% | 161% | 101% | 30% |
| Mathematics and computer sciences | 54% | 168% | 111% | 37% |
| Medicine | 53% | 209% | 143% | 41% |
| Physics | 43% | 162% | 104% | 33% |

*Table 7: Coefficients of variation for SS within and between Italian universities.*

The same considerations hold true for SS. Giving the example of Civil engineering and architecture, we see that all the universities active in the UDA, except one, show variability that is greater than that between universities (Figure 6).

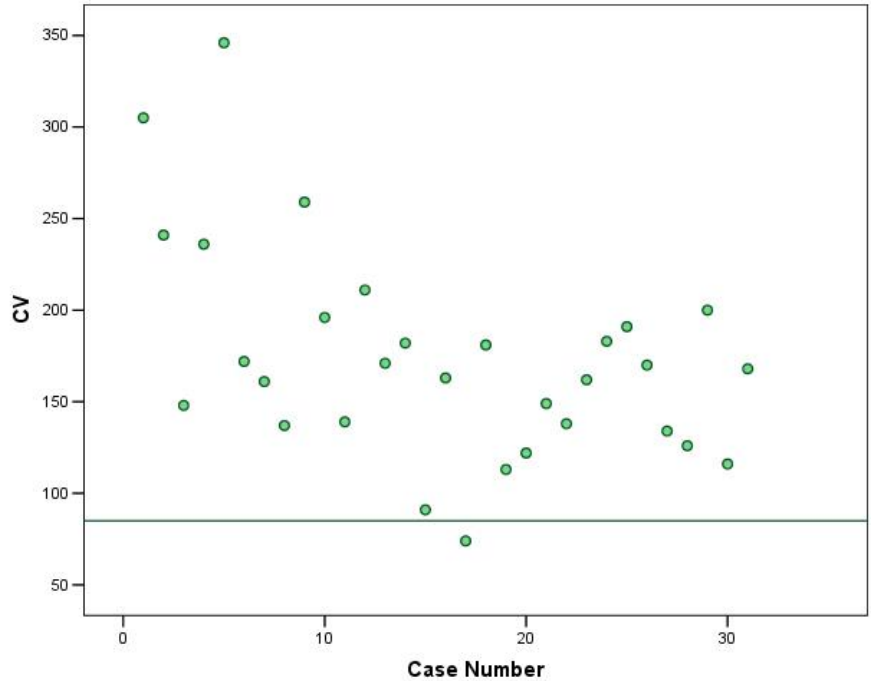

*Figure 6: Coefficients of variation per SS within (points) and between (line) Italian universities.*

The individual universities hold values of between 74% and 346%, while variability among them is 85%. The Biology UDA shows one university with dispersion of performance lower than that between universities. However, all other UDAs conform to the observation that variability of performance within universities is always greater than that between (Table 8). The UDA with the lowest and greatest median values of dispersion for SS are, respectively, Chemistry (125%) and Medicine (190%).

|  | $C_V$ within |  |  | $C_V$ between |
|---|---|---|---|---|
| UDA | min | max | median |  |
| Agricultural and veterinary sciences | 81% | 245% | 159% | 44% |
| Biology | 54% | 297% | 159% | 58% |
| Chemistry | 71% | 303% | 125% | 33% |
| Civil engineering and architecture | 74% | 346% | 160% | 85% |
| Earth sciences | 61% | 236% | 129% | 58% |
| Industrial and Information engineering | 74% | 267% | 161% | 37% |
| Mathematics and computer sciences | 87% | 295% | 159% | 55% |
| Medicine | 68% | 313% | 190% | 59% |
| Physics | 51% | 204% | 136% | 45% |

*Table 8: Comparison of coefficients of variation for SS within and between universities.*

The results described thus show that the variability of performance between universities is less than variability of performance of individual scientists within each Italian university, confirming expectations, given the non-competitive environment. In the Italian university system there is a flattening of the average level of performance of universities, due to the high concentration of performance of researchers within each university.

**4.4 Degree of concentration and average level of performance**

From the preceding analyses it emerges that degree of concentration of performance within Italian universities is generally high or very high and greater than that between them. In this section we test whether the values of concentration of each university are or are not correlated with the values for performance. We test at the levels of the single UDA and globally.

Applying the Spearman non-parametric coefficient of correlation at the level of UDA, between P and the relative degree of concentration, we obtain significant correlations in 6 UDAs out of 9 (Table 9). Apart from Chemistry, correlation is always negative. The highest value of correlation ($\rho$ = -0.606, p-value <0.01) is seen in Medicine. In this UDA highly productive universities show on average a very low concentration of performance at individual level. This can be due to the specificity of the "research production function" in this UDA. In medicine, team-work is practically the norm, while in other disciplines, such as mathematics, is much less so. A large quota of collaborations occur mainly intramuros, so that if a team is highly productive on average, all of its members are highly productive as well, and viceversa. Anyway, concerning SS, the correlation is significant in 4 UDAs. In this case again, the only positive correlation occurs in Chemistry, but with p-value >0.10, therefore not statistically significant.

| UDA | P vs Gini P | SS vs Gini SS |
|---|---|---|
| Agricultural and veterinary sciences | -0.262 | -0.113 |
| Biology | -0.393*** | -0.244** |
| Chemistry | +0.110 | +0.191 |
| Civil Engineering and architecture | -0.398** | -0.113 |
| Earth sciences | -0.287* | -0.202 |
| Industrial and Information engineering | -0.106 | -0.147 |
| Mathematics and computer sciences | -0.495*** | -0.223* |
| Medicine | -0.606*** | -0.618*** |
| Physics | -0.365*** | -0.330** |

*Table 9: Spearman correlation between P, SS and Gini coefficients*
*Significance level: \*p-value <0.10, \*\*p-value<0.05,\*\*\*p-value<0.01.*

Testing for global correlation (without distinguishing by UDA), we obtain a significant and negative correlation for both indicators of performance. The highest correlation is obtained for the correlation of P and the relative degree of concentration ($\rho$ = -0.293, p-value < 0.01), while that for SS is lower ($\rho$ = -0.182, p-value < 0.01).

The negative relationship between performance and relative concentration, even though not high in some disciplines, is confirmed by the comparison of concentration between the best and worst universities in terms of performance. Two subgroups were created for this test: the first consisting of the universities in the top 10% for performance in their sectors, for every disciplinary area, and the second group consisting of the bottom 10%.

The analysis shows that the average level of concentration for both measures of performance is less within the top 10% subgroup than in the bottom 10% group (Table 10). There is thus less dispersion of performance within the best universities, although it is still high. The only exception is the Chemistry UDA, where the situation is completely reversed (note that the coefficients of correlation were positive). This result is not completely surprising: unlike the other UDAs, Chemistry has a very low average value of unproductive researchers, at 6%, which dictates a greater equidistribution of productivity among its researchers. Also, the Earth sciences UDA shows, limited to SS, a concentration of performance in the top 10% that is greater than in the bottom 10%. Chemistry, for P, and Earth sciences, for SS, show the lowest levels of concentration among all the UDAs.

|  | Average Gini P | | Average Gini SS | |
| --- | --- | --- | --- | --- |
| UDA | Top 10% | Bottom 10% | Top 10% | Bottom 10% |
| Agricultural and veterinary sciences | 0.467 | 0.475 | 0.591 | 0.650 |
| Biology | 0.491 | 0.512 | 0.540 | 0.559 |
| Chemistry | 0.415 | 0.394 | 0.516 | 0.437 |
| Civil Engineering and architecture | 0.505 | 0.648 | 0.583 | 0.700 |
| Earth sciences | 0.479 | 0.560 | 0.474 | 0.341 |
| Industrial and information engineering | 0.447 | 0.451 | 0.582 | 0.673 |
| Mathematics and computer sciences | 0.446 | 0.536 | 0.598 | 0.634 |
| Medicine | 0.435 | 0.623 | 0.578 | 0.710 |
| Physics | 0.452 | 0.525 | 0.522 | 0.589 |

*Table 10: A comparison of Gini coefficients for P and SS between top 10% and bottom 10% universities in each UDA*

The results are confirmed by the NPC test, a combination of non-parametric tests for comparison between groups, stratified by UDA (Pesarin, 2001). The test is for $GINI_{top10\%} = GINI_{bottom10\%}$, against the alternative hypothesis that $GINI_{top10\%} < GINI_{bottom10\%}$, both globally and within each UDA. The combined Fisher test shows significant differences between the two groups, both for P (p-value = 0.005 < 0.01) and for SS (p-value = 0.034 < 0.05) (last line of Table 11): the concentration of the top 10% is on average less than the concentration of the bottom 10%.

Examining the different UDAs, significant differences in the levels of concentration of P are seen only in Medicine and in Civil engineering and architecture (both with p-value < 0.05), and in Mathematics and computer sciences (p-value < 0.10). For levels of concentration of SS, significant differences appear in Medicine (p-value < 0.05) and in the engineering areas (Table 11).

|  | p-value | |
| --- | --- | --- |
| UDA | Gini P | Gini SS |
| Agricultural and veterinary sciences | 0.454 | 0.448 |
| Biology | 0.372 | 0.386 |
| Chemistry | 0.640 | 0.960 |
| Civil Engineering and architecture | 0.014 | 0.074 |
| Earth sciences | 0.167 | 0.790 |
| Industrial and Information engineering | 0.485 | 0.083 |
| Mathematics and computer sciences | 0.089 | 0.247 |
| Medicine | 0.021 | 0.013 |
| Physics | 0.171 | 0.226 |
| *Combined test using the Fisher combination* | *0.005* | *0.034* |

*Table 11: NPC test, stratified by UDA and combined*

## 5. Conclusions

The Italian higher education system presents levels of concentration of performance within universities that are high or very high, and greater than that between them. This situation should be typical of non-competitive university systems, which do not favor the search for a sustainable competitive advantage, with the resulting formation of outstanding institutions, capable of attracting, developing and retaining the top scientists from the nation and from abroad. Dispersion of performance may thus reveal an important indicator of intensity of competitiveness in higher education systems. The implications at the level of national and institutional policies can not be ignored. In non-competitive systems like the Italian one, the allocation of a part of public resources to individual universities on the basis of results from national exercises of evaluation, which compare the average performance of the universities, could result as having little effective use. If the intent of the policy-maker is that of optimizing the socio-economic impact of public financing for research, through allocating greater resources to universities that have a higher average efficiency, the high dispersion of performance within universities leaves open the risk that the macroeconomic objective will not be reached, unless the universities internally allocate the resources according to similar principles of merit. This risk is much less in competitive systems, where the variability of performance within is much less than between universities. In this case, allocation of resources to excellent universities is more or less equivalent to allocating resources to the best scientists. In non-competitive systems, as in the Italian case, it is unlikely that the universities will have the capacity, even if they have the desire, to identify the best research products, and thus the best scientists. Abramo et al. (2009) have in fact demonstrated the limited effectiveness of Italian universities in selecting their best research products for the last national exercise for evaluation of research. An immediate policy recommendation is thus that national evaluation exercises be conducted at the level of the individual, at least in the hard sciences where bibliometric techniques make this possible, in order to inform universities' internal selective funding.

Even if there is efficient allocation at the level of individuals, the authors remain skeptical that when the share of financing based on merit is very small (as in the Italian case) and without an accompanying incentive systems linking salaries to merit, this would then lead to significant increments in production efficiency or to efficient selection that is typical of competitive systems, replacing deeply-rooted practices of favoritism. In the current economy of knowledge, in order that this be realized with the urgency that the current global competitive challenges require, a more daring policy recommendation by the authors is that of favoring the budding of spin-off universities, possibly with balanced regional distribution, and with in-migration of only the top scientists from the national public research system. With a very low investment, for the marginal costs of transfers, it would be possible to quickly create the type of top universities that competitive systems in other national contexts have produced over the span of decades: universities capable of attracting the best scientists, students and enterprises to the region, and leading to much greater economic benefit than the present universities, with their high dispersion of internal performance, are able to produce.

A further consideration concerns the line of research which focuses on comparisons of research performance of national systems (such as May, 1997; King, 2004; Dosi et al., 2006). These comparisons, which among other factors usually omit to carry out field-standardizations, are based on the measure of central tendency, without delving into how data is dispersed. Two higher education systems with the same average research performance could have impacts on national socio-economic development that are completely different, all others equal, according to the dispersion of performance within each institution. A measure of dispersion is thus just as important as that of central tendency in international comparison of research systems. Finally, international rankings of universities, such as Thes[14], which also base their results on measurements from perceptions of the universities' reputations, are inappropriate in non-competitive national systems. For these, the dispersion of performance within the universities is so high that there is no emergence of the type of top university seen in competitive systems, thus making it difficult to perceive strong differences in performance between the universities, which in fact, as this study demonstrates, do not exist.

It is the hope of the authors that colleagues from other national contexts carry out similar studies and provide data on the dispersion of performance in other countries as well. Further analyses, on this stream of research, may concern the impact of structural variables such as size, geographical location, etc. on the concentration of performance. Furthermore, time-series analysis of levels of performance concentration may reveal the effectiveness of policies aimed at stimulating more competition in higher education systems.

---

[14] See http://www.timeshighereducation.co.uk/hybrid.asp?typeCode=431